# Assembling nanostructures from DNA using a composite nanotweezers with a shape memory effect

1st Andrey P. Orlov
*Kotelnikov Institute of Radioengineering and Electronics of RAS*
*Institute of Nanotechnology of Microelectronics of RAS*
Moscow, Russia
e-mail: andreyorlov@mail.ru.

2nd Anatoly M. Smolovich
*Kotelnikov Institute of Radioengineering and Electronics of RAS*
Moscow, Russia
e-mail: asmolovich@petersmol.ru.

3d Nikolay A. Barinov
*Federal Research and Clinical Center of Physical-Chemical Medicine*
Moscow, Russia
e-mail: nik_mipt@mail.ru

4th Aleksei V. Frolov
*Kotelnikov Institute of Radioengineering and Electronics of RAS*
Moscow, Russia
e-mail: fralek@mail.ru.

5th Peter V. Lega
*Kotelnikov Institute of Radioengineering and Electronics of RAS*
Moscow, Russia
lega_peter@list.ru.

6th Dmitry V. Klinov
*Federal Research and Clinical Center of Physical-Chemical Medicine*
Moscow, Russia
e-mail: klinov.dmitry@mail.ru

7th Victor V. Koledov
*Kotelnikov Institute of Radioengineering and Electronics of RAS*
Moscow, Russia
e-mail: victor_koledov@mail.ru.

*Abstract*— **The article demonstrates a technique for fabricating a structure with the inclusion of suspended DNA threads and manipulating them using composite nanotweezers with shape memory effect. This technique could be suitable for stretching of nanothin DNA-like conductive threads and for measuring their electrical conductivity, including the I-V characteristic directly in the electron microscope chamber, where the nanotweezers provide a two-sided clamping of the DNA tip, giving a stable nanocontact to the DNA bundle. Such contact, as a part of 1D nanostructure, is more reliable during manipulations with nanothreads than traditional measurements when a nanothread is touched by a thin needle, for example, in a scanning tunnel microscope.**

## I. Introduction

The study of transport properties of DNA has been carried out for a long time and is still of great interest due to the uniqueness of this molecule. However, the conductivity of the canonical double stranded DNA is very low. Recently great progress was achieved in the metallization of DNA [1-3]. After binding of metal ions to a DNA template the conductivity of molecule increases significantly. Such conductive nanowires could be used for mechanical and electrical measurements during stretching.

Our purpose was to develop a system, which allows to capture single DNA wire and to manipulate it. We consider that the most suitable tool to to perform this task is nanotweezers with shape memory effect (SME). Functional materials based on Ni can change their forms or dimensions in response to an external impact, such as a temperature change, magnetic, or electric field. Such materials are very important, particularly in constructing NEMS, since conventional mechanical systems are unsuitable for small dimensions. Under the external action, the alloys with SME can change the shape of the active element. However, the reversibility of this change can be achieved only through a special nontechnological process called "training". In our case using nanotweezers with SME we can fix molecules between conductive contact pads and hold them strong during the stretching.

To accomplish multiple reversible deformations of the nanotweezers with SME, a composite scheme was used [4]. In order to manufacture the actuator, rapidly-quenched ribbons of $Ti_2NiCu$ alloy with SME, obtained by the method of superfast melting on a rotating copper disk, were used. This procedure of manufacturing composite nanotweezers with SME is further described in [5, 6]. Fig. 1 shows the gripping part of the nanotweezers made from $Ti_2NiCu$ alloy in an open (martensitic) state with a working gap of 200 nm.

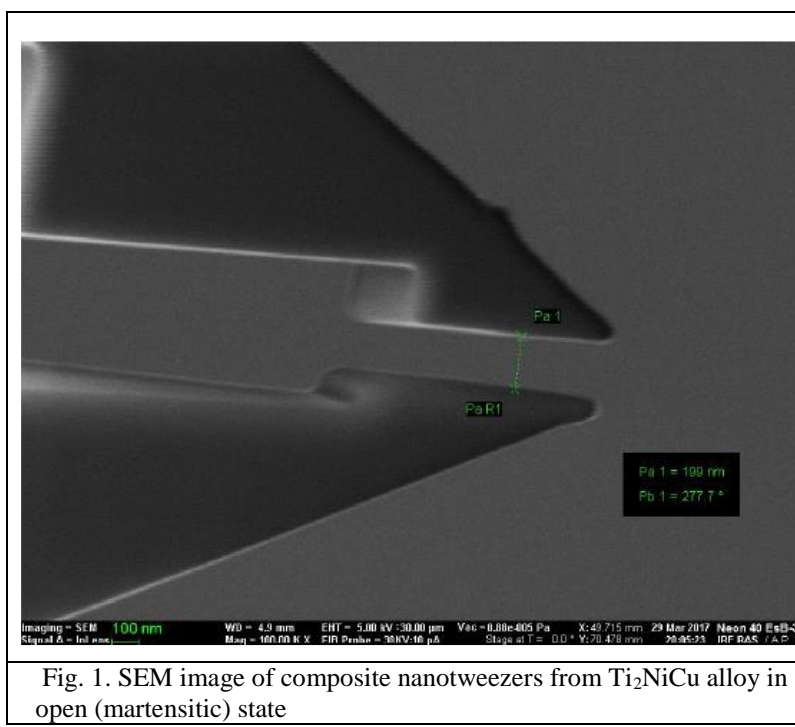

Fig. 1. SEM image of composite nanotweezers from $Ti_2NiCu$ alloy in open (martensitic) state

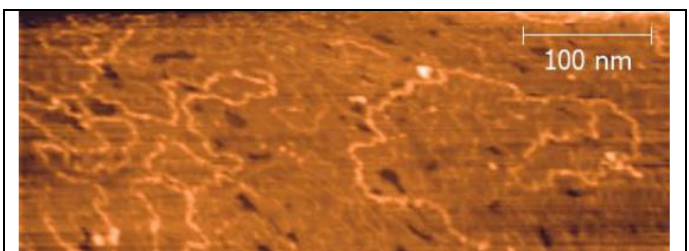

Fig. 2. High-resolution atomic force microscopy of single DNA molecules on graphite. Colors in topography images represent the height variation Δh = 5 nm.

We began our manipulation attempts with samples of DNA placed on graphene and graphite substrate. Recently, we obtained hybrid structures of DNA-graphene and visualized single DNA molecules using atomic force microscope (AFM) (fig.2). The possibility of controlled application of DNA to graphene films, obtained by mechanical splitting of graphite on a substrate with a sub-layer of epoxy glue was demonstrated. There is a great interest in these structures [7-11] due to the possibility of using such structures as biosensors, in particular, for medical diagnostics. Graphene can also be used as a storage medium for DNA, for hybridization, for address assembly and for the interaction of complementary chains [12]. It is also of interest to cover graphene surface by metallized DNA, which is essentially a molecular wire [1, 13]. When applied, the conductivity of both graphene and the DNA molecules may change.

## II. OBTAINING OF SUSPENDED DNA THREADS

Our first attempt was to tear DNA off the graphene sheet using maximally sharp nanotweezers, but this technique proved to be unsuitable because of the extremely low thickness of DNA molecules (below 1 nm) and of their strong attraction to the substrate.

Then we elaborated a technique of obtaining suspended DNA molecules. In this case, DNA bundles were placed on membranes between thin cuts, and we could use our nanotweezers to manipulate DNA in these cuts to assemble conductive nanostructures from DNA, as shown in Fig. 3. [14]. According to the scheme, DNA molecules are directly connected to the current-voltage (I-V) measurement circuit through the nanotweezers and the substrate. Such measurements could be interesting for studying the electronic transport in metallized DNA threads.

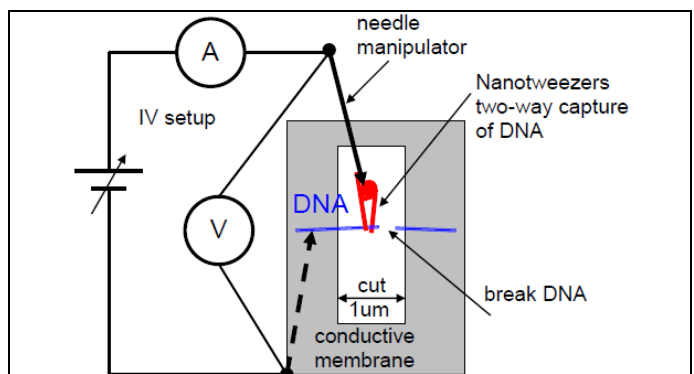

Fig. 3. Scheme for measuring the electrical conduction of DNA molecules using composite nanotweezers with SME

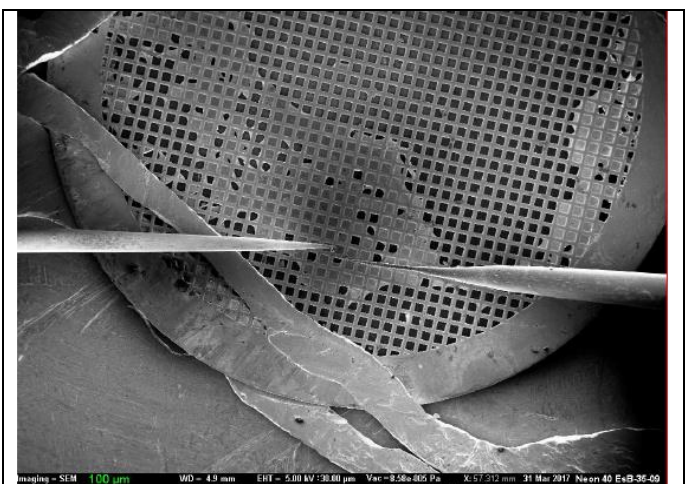

Fig. 4. Mesh covered with thin membrane.

For the preparation of suspended DNA samples, a membrane of nitrocellulose was first used as a carrier for DNA molecules. The thickness of the membrane was about 50 nm. The membrane was placed on a copper mesh for a transmission microscope with square cells with a side of 50 μm, or alternatively the membrane was placed on a gap between two silicon wafers. To increase adhesive properties of the membrane, a layer of conducting carbon with a thickness of 10 nm was deposited (fig. 4). Then, using a focused ion beam (FIB), rectangular narrow (~ 1 μm) cuts 10-20 μm in length were etched in the membrane (fig. 5).

To apply DNA molecules on a graphene or a carbon membrane, a similar technique for applying DNA molecules to graphite [14-17] was used. On the surface, 100 μl of 0.1% graphite modifier solution GM $(CH_2)n-(NCH_2CO)m-NH_2$, (manufactured by Nanotyuning, Chernogolovka, Russia), was exposed for 1 minute, then GM was removed in a compressed nitrogen and the substrate was dried. Molecules of duplex DNA from bacteriophage lambda (Escherichia virus Lambda) at a concentration of 1 μg / ml were applied from a solution of 10 mM Tris-HCl (pH 7.6), 1 mM EDTA, to the surface of the modifier for 1 min, then a drop of DNA was removed into a compressed nitrogen stream. In the case of carbon coated membrane from nitrocellulose, a drop of DNA was removed by drying without blowing, with a piece of filter paper.

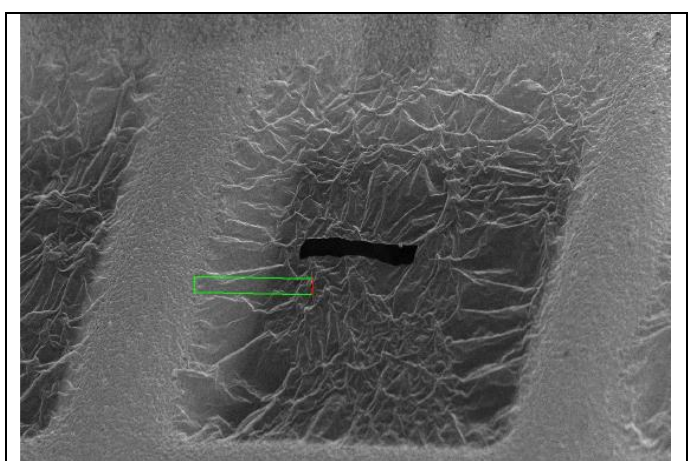

Fig. 5. Separate cell with a rectangular cut in the membrane.

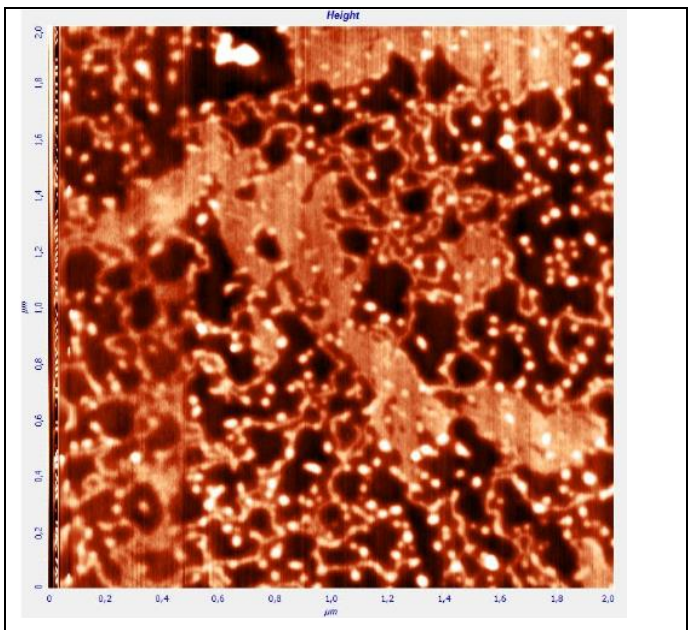

Fig. 6. High-resolution atomic force microscopy of DNA molecules on a membrane applied by drying without blowing (in the figure, the DNA is mixed with impurities). Colors in topography images represent the height variation $\Delta h = 10$ nm

The monolayer of the modifier is a lamellar structure epitaxially crystallized by intermolecular hydrogen bonds on the graphite surface. It is necessary to attach DNA molecules to the surface in the expanded state, because DNA weakly interacts with pure graphite, it twists, coagulates and shifts as the meniscus of the drop passes through it. If the drop is not blown off, but only dried, the impurities in solution in mass fractions greater than $10^{-6}$ will cover the surface with a rough layer comparable to DNA thickness ~ 1nm, making it difficult to identify DNA by the AFM method. The use of filter paper to remove the residuals of DNA solution helps to reduce the amount of impurities on the surface (fig.6).

The control was carried out using an atomic force microscope (AFM) [18] (fig. 6). The relief of the samples was measured in a semi-contact resonant mode at the NT-MDT Integra Prima research complex using the NOVA 1.1 control program. The ultra-sharp cantilevers (carbon nanowhiskers with a curvature radius of several nanometers grown at tips of commercially available silicon cantilevers, with a resonant frequency in the range of 190-325 kHz, and an angle at its apex <22°, produced by Nanotyuning, Russia) were used. The amplitude of free oscillations of the cantilever in the air lay in the range of 1-10 nm, the automatically maintained amplitude of the cantilever oscillations in the state approaching the surface (SetPoint parameter) was set at 60-70% of the amplitude of the free vibrations of the cantilever in the air.

The width of the cuts varied from parts of microns to a few microns. In addition to rectangular cuts, more complex shapes were used. A thin gold layer (1-2 nm) was deposited on the membrane so that the remaining DNA molecules could be observed in the electron microscope. Observations showed that the molecules of the set of DNA were stuck together in a kind of bundles tens of nanometers thick, some of which fell on the incisions in the membrane, Fig. 7.

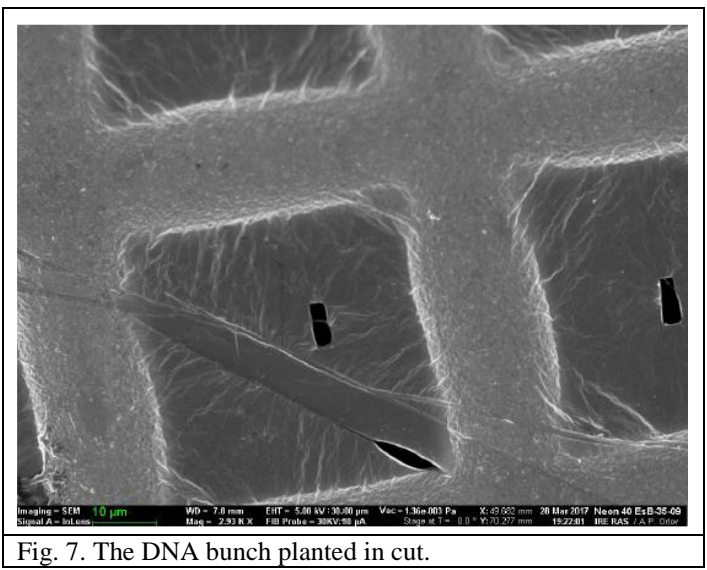

Fig. 7. The DNA bunch planted in cut.

## II. SINGLE DNA THREADS MANIPULATION AND APPLICATION NANOSTRUCTURES FOR MEASUREMENTS

The main stages of the manipulation with a bundle of DNA molecules using composite nanotweezers and of the production of a two-contact structure for measuring the I-V characteristic in an electron microscope chamber CrossBeam Neon40 EsB (Carl Zeiss) with manipulators Kleindiek are shown in Fig. 8 – 10.

The device (nanotweezers), including a tungsten needle, a heating unit (silicon chip diode), and the nanotweezers, was mounted and connected to an electronic control circuit. The tungsten needle is compatible with commercial manipulators Kleindiek. Operations on manipulating with DNA threads were held using such manipulators.

Fig. 8 shows the process of approximation of the nanotweezers to the thread. When the thread is between the jaws of the nanotweezers, a current is fed, and with the help of the diode, the composite nanotweezers are heated and transformed into the austenitic state. This causes compression of the jaws and, consequently, the capture of the bunch, Fig. 8. After that, the nanotweezers with the bunch of DNA could be moved to stretch the bunch or to tear it.

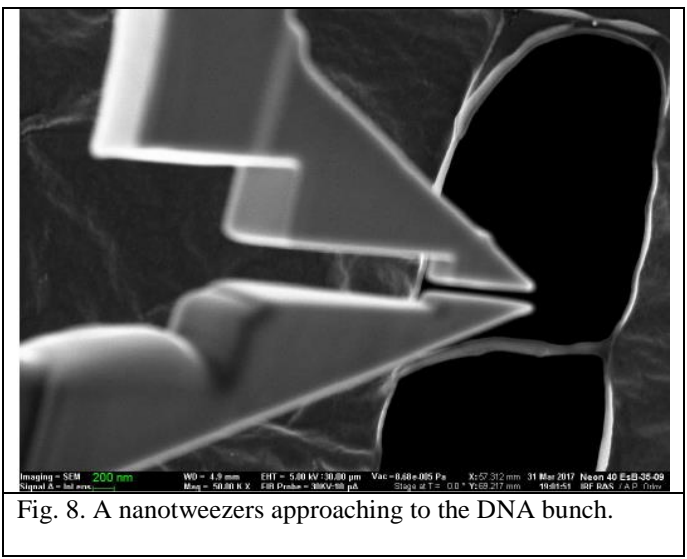

Fig. 8. A nanotweezers approaching to the DNA bunch.

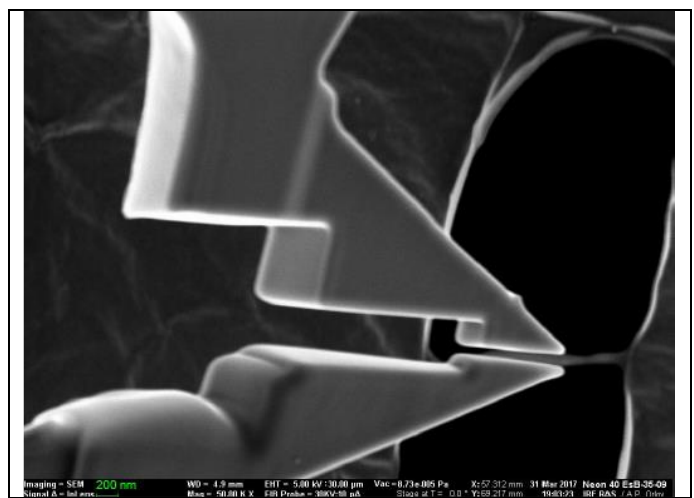
Fig 9. Capturing DNA bunch by composite nanotweezers.

Usually, it is impossible to press a single needle with one end of the suspended DNA and, moreover, to get a reliable electrical contact, but using nanotweezers it is possible to grab DNA without problems, squeezing it on both sides. Moving tweezers can not only stretch the DNA, but also separate the molecule from the edge of the membrane, as shown in Fig. 10. This allows it to be fixed between two electrical contacts, according to the scheme in Fig 1, where one contact is a conductive membrane, the other one is conductive nanotweezers. Such incorporation of DNA into the electrical circuit for I-V measurements is of scientific interest for studying the electrical conductivity of chains of DNA molecules. If the nanostructures obtained from DNA nanowires are measurable, the methods that were used for the study of carbon nanotubes (CNT) and Bi-nanowires could be applied. [19, 20].

## IV. CONCLUSION

Using the nanotweezers made from material with a shape memory effect developed by the authors and advanced technique for fixing DNA molecules on a membrane with microcuts, a nanostructure with a metalized (conducting) DNA molecule (bundles of several chains) was assembled and prepared for studying of its electrotransport properties in a chamber of a scanning electron microscope.

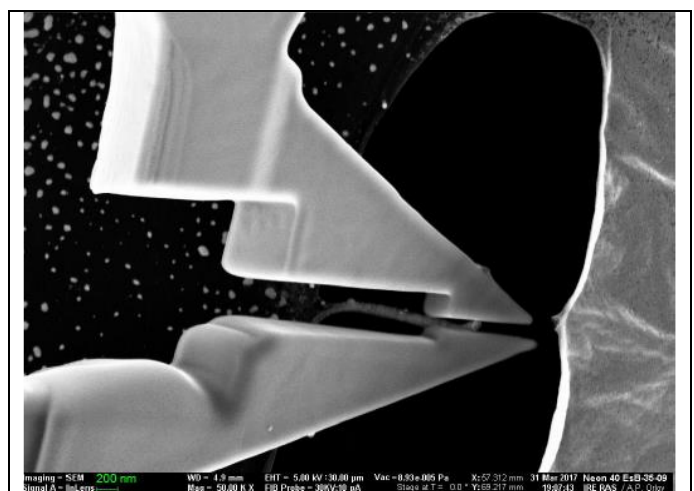
Fig. 10. Break out of the DNA bunch by composite nanotweezers.

AKNOWLEDGEMENTS
The work is supported by RSF Grant No 17-19-01748